\documentclass[prl,twocolumn,draft,amsmath,showpacs]{revtex4}
\usepackage{graphics}

\begin{document}
    
\bibliographystyle{prsty}
\input epsf

\title {Anisotropic field dependence of the magnetic transition in Cu$_2$Te$_2$O$_5$Br$_2$}

\author {A.V. Sologubenko, R. Dell'Amore and  H.R. Ott}
\affiliation{Laboratorium f\"ur Festk\"orperphysik, ETH H\"onggerberg,
CH-8093 Z\"urich, Switzerland}
\author {P. Millet}
\affiliation{Centre d'Elaboration de Mat\'eriaux et d'Etudes Structurales, 
CEMES/CNRS, F-31062 Toulouse, France}

\date{\today}

\begin{abstract}
In this paper, we present the results of measurements of the thermal conductivity of  
Cu$_2$Te$_2$O$_5$Br$_2$,
a compound where tetrahedra of  Cu$^{2+}$ ions carrying $S=1/2$ spins  form chains 
along the $c$-axis 
of the tetragonal crystal structure. 
The thermal conductivity $\kappa$ was measured along both the $c$- and 
the a-direction  as a function of temperature between 3 and 
300~K  and in external magnetic fields $H$ up to 69~kOe, oriented both parallel and perpendicular to the $c$-axis. Distinct features of $\kappa(T)$  were 
observed in the vicinity of $T_{N}=11.4$~K in zero magnetic field.
These features are unaltered in external fields which are 
parallel to the $c$-axis, but are more pronounced when a field is applied perpendicularly to the $c$-axis. The transition temperature increases upon enhancing the external field, but only if the field is oriented along the $a$-axis.
\end{abstract}
\maketitle

\section{Introduction}
Thermal transport in low-dimensional quantum spin systems has recently been investigated in detail, both experimentally and theoretically. 
Considerable progress has been made in the theoretical understanding of heat transport in idealized one-dimensional (1D) and two-dimensional (2D) model spin systems.
For 1D systems that are dominated by 
antiferromagnetic (AFM) couplings of the spins, the integrability of the corresponding model Hamiltonians leads to interesting and nontrivial results for various transport properties, including spin-, charge-, and energy-transport \cite{Castella95,Zotos97,Saito96,Narozhny98,Naef98,Kluemper02,Alvarez02_Ano,HeidrichMeisner02,Saito03}. 
For the description of heat transport in real materials, more realistic models, considering perturbations such as spin-lattice coupling, defects and three-dimensional interactions have to be considered. Only a small amount of theoretical work along these lines is available in the literature. For example, the spin-phonon coupling in Heisenberg AFM $S=1/2$ spin chains has been considered and calculations including the interaction of spins with defects in AFM $S=1/2$ spin chains and ladders were made \cite{Shimshoni03,Orignac03}.     

Experimental investigations treating quasi-2D spin systems concentrated on 
measurements of thermal transport in layered cuprates and vanadates \cite{Nakamura91,Cohn95,Hess03,Hofmann03,Sun03,Sales02} and in the Shastry-Sutherland spin-lattice compound SrCu$_2$(BO$_3$)$_2$ \cite{Vasilev01,Hofmann01,Kudo01_The}. 
For quasi-1D systems, experimental results were reported for $S=1/2$ spin ladders \cite{Sologubenko00_lad,Hess01} and $S=1/2$  Heisenberg spin chains \cite{Sologubenko01,Sologubenko03_Uni,Sologubenko03_Dif,Markina03}, including the inorganic spin-Peierls compound CuGeO$_3$ \cite{Vasilev97, Ando98, Salce98,
Vasilev98, Takeya00, Takeya00a, Hofmann02,
Takeya01}. The common feature of all these compounds is that the heat transport is dominated by phonons, except along the directions of strong spin-spin interactions, i.e., along the chains in 1D systems and in the plains in 2D systems.
For these cases, significant heat transport carried by spin excitations is observed in limited  temperature intervals. In most materials, however, the spin system simply acts as a source of phonon scattering and cannot be regarded  as a channel of significant energy transport.     

The  material studied in the present work may be viewed as a quasi-zero-dimensional spin system. The essential structural subunits are weakly interacting  spin tetrahedra. The ground state and the excited states of noninteracting spin tetrahedra are well understood \cite{Johnsson00,Lemmens01}. Various anisotropic and frustrated interactions between spin tetrahedra, even if relatively weak, lead to interesting and non-trivial ground states and quantum phase transitions. An acceptable physical realization of the spin-tetrahedra model was recently found in compounds of the type Cu$_2$Te$_2$O$_5X_2$ with $X$=Cl or Br \cite{Johnsson00,Lemmens01}, for which
the tetragonal crystal structure of Cu$_2$Te$_2$O$_5X_2$ is formed by  distorted tetrahedra of Cu$^{2+}$ ions aligned along the $c$-axis \cite{Johnsson00}. Two types of AFM bonds within the tetrahedra are associated with exchange integrals $J_1$ and $J_2$. The magnetic susceptibility $\chi(T)$ exhibits a peak and subsequently decreases exponentially  with decreasing $T$ at low temperatures. This suggests that the spins are dimerized and the corresponding energy gap $\Delta/k_B$, separating the ground state from excited states, is about 40~K \cite{Johnsson00,Lemmens01}. 
The analysis of the susceptibility data, assuming that $r \equiv J_2/J_1  =1$, results in $J_1/k_B=J_2/k_B= 38.5$ and 43~K for $X=$ Cl and Br, respectively \cite{Johnsson00,Lemmens01}. 
A mean-field (MF) type analysis of results of Raman scattering measurements on the Br-compound suggests that $r=0.66$, and $J_1/k_B=47 {\rm ~K}$.  The inter-tetrahedral coupling parameter $J_c = 0.85 J_1$ \cite{Gros03} leads to phase transitions to an AFM ordered state at $T_N = 18.2$ and 11.4~K for $X=$ Cl and Br, respectively \cite{Lemmens01}. 
The application of an external magnetic field $H$ reduces
$T_N$ for  Cu$_2$Te$_2$O$_5$Cl$_2$. For Cu$_2$Te$_2$O$_5$Br$_2$, however, an unusual increase of the  transition temperature with increasing $H$ was observed. It was argued \cite{Lemmens01,Gros03} that the latter anomalous behavior is caused by the vicinity of a quantum critical transition, which is expected if $J_c=0.75 J_1$. 
Several theoretical models have since been put forward to explain the magnetically ordered state and the related excitation spectrum  of Cu$_2$Te$_2$O$_5$Br$_2$. The essential inputs were based on invoking, e.g., anisotropic inter-tetrahedral and Dzyaloshinsky-Moriya-type interactions \cite{Brenig01,Totsuka02,Jensen03,Kotov04cm}, but
the nature of the low-temperature phase still remains largely unexplained. 
Very recent neutron diffraction experiments \cite{Zaharko04cm} were interpreted as to indicate the formation of an incommensurate long-range magnetic order for both $X$=Cl and Br. 
  
Results on the temperature dependences of the thermal conductivity $\kappa(T)$ of Cu$_2$Te$_2$O$_5X_2$ for both $X$=Br and $X$=Cl in zero magnetic field were recently reported in Ref. \cite{Prester04}. The authors observed a strong anomaly in $\kappa(T)$ near $T_N$ for Cu$_2$Te$_2$O$_5$Cl$_2$ and attributed it to an unexpectedly large spin-lattice coupling in this compound. In contrast, no anomaly was observed for the isomorphic Cu$_2$Te$_2$O$_5$Br$_2$, which was ascribed to an intrinsically weak spin-lattice coupling in this material.   
The present work includes measurements of the low-temperature thermal conductivity $\kappa(T,H)$ of Cu$_2$Te$_2$O$_5$Br$_2$ with a special emphasis on investigating the influence of external magnetic fields $H$ with different orientations. In the vicinity of the ordering transition, we observe a pronounced anomaly of $\kappa(T)$, which is remarkably sensitive to the strength and the orientation of the external magnetic field. 
The main result is the observation of an anomalous increase of $T_N$ with increasing $H$, but only if the field orientation is perpendicular to the $c$-axis, at least up to 6~T. The implications of this observation are discussed in view of recent theoretical suggestions for the cause of the magnetic ordering transition in  Cu$_2$Te$_2$O$_5$Br$_2$.  

\section{Samples and experiment}

The samples for this investigation were cut from a large single crystal of 
Cu$_2$Te$_2$O$_5$Br$_2$, grown as described in 
Ref. \cite{Johnsson00}. Two bar-shaped samples with approximate
dimensions of $0.5\times 0.5\times 2$~mm$^{3}$ were cut in such a manner 
that the longest direction was, for one sample, along the $c$-direction, and perpendicular to the $c$-direction for the other specimen. 
The thermal conductivity was measured  in the 
temperature region between 2 and 300~K by using the standard method of uniaxial 
heat flow as described in Ref. \cite{Sologubenko03_Dif}.  The 
magnetic fields were oriented along either the $c$- or 
the $a$-axis of the crystal structure. Complementary measurements of the magnetic 
susceptibility were made with a commercial SQUID magnetometer at temperatures between 2 and 300~K.

\section{Results}

\begin{figure}
    \begin{center}
  \epsfxsize=1\columnwidth \epsfbox {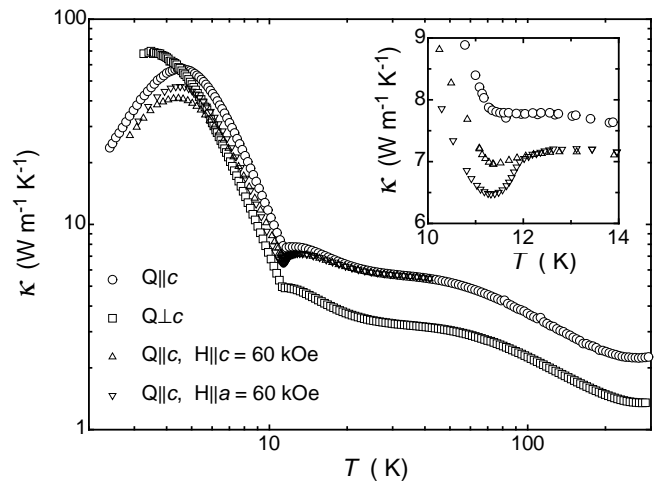}
   \caption{
  Thermal conductivity vs. temperature of Cu$_2$Te$_2$O$_5$Br$_2$ along and perpendicular to the $c$-axis in zero magnetic field and in $H=60 {\rm ~kOe}$. The inset emphasizes $\kappa(T)$ for the heat flow parallel to the $c$-axis in the vicinity of the magnetic ordering transition.  
  }
	\label{KvsT_H0}
    \end{center}
\end{figure}

The temperature dependences of the thermal conductivities, $\kappa(T)$, 
along two crystallographic orientations in zero magnetic field and in $H=60$~kOe are shown in Fig. \ref{KvsT_H0}.
The general features of $\kappa(T)$ along the two heat flux directions are essentially the same for $H=0$, especially above 
approximately 7~K, where the data for the two samples differ  by practically 
a constant factor, such that  $\kappa_{\parallel c}/\kappa_{\perp c} \approx 1.6$. 
At lower 
temperatures, this ratio is gradually reduced to  about 0.6 at 3~K.  
Each $\kappa(T)$ curve in Fig. \ref{KvsT_H0} exhibits 
a maximum between 3.5 and 4.5~K and a distinct feature around 
$T_{N}=11.4 {\rm ~K}$.  
This type of low-temperature maximum of $\kappa(T)$  is typical for insulators and,
with increasing temperature, 
reflects the gradual change from the dominant boundary scattering to enhanced phonon-phonon scattering of the itinerant lattice excitations.  
A sharp feature of $\kappa(T)$ is usually related to some kind of phase transition, in the present case to magnetic ordering. 
Applying an external magnetic field well above $T_{N}$ leads to only a slight decrease  
of $\kappa$, almost independent of the field 
orientation. However, a significant and $H$-orientation dependent 
reduction of the thermal conductivity  by magnetic field is observed in the 
vicinity and below $T_N$ (see the inset in Fig. \ref{KvsT_H0}). 

\begin{figure}[t]
 \begin{center}
  \epsfxsize=1\columnwidth \epsfbox {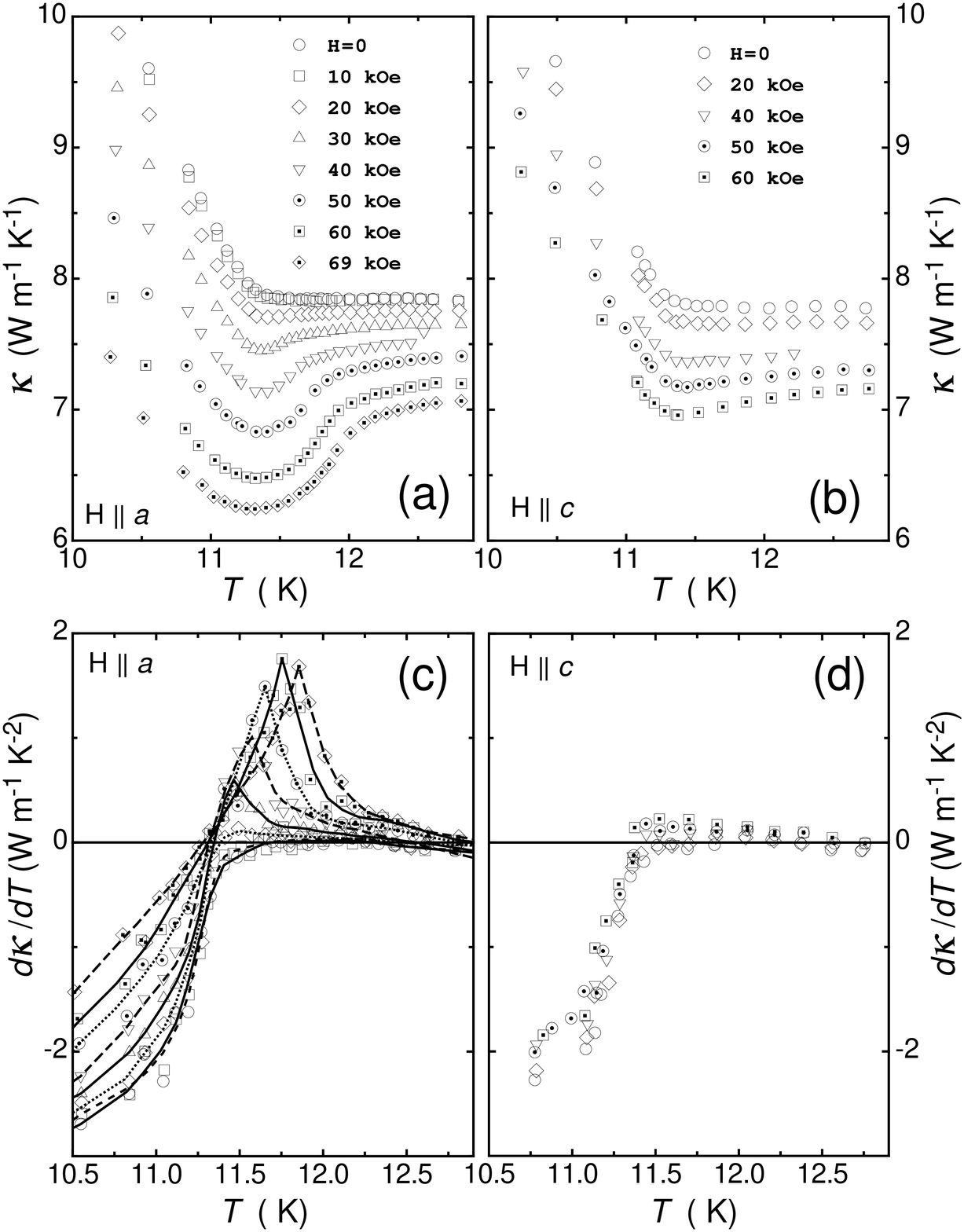}
   \caption{
  (a,b) $\kappa(T)$ of Cu$_2$Te$_2$O$_5$Br$_2$ along the the $c$-direction in the vicinity of 
  $T_{N}$ in different magnetic fields which are oriented parallel to the $a$- and $c$-axes, 
  respectively. (c,d) The corresponding temperature derivatives 
  $\partial\kappa /\partial T$ vs. $T$ for the same field orientations as in (a,b).  The lines in (c) are guides for the eye. 
  }
\label{KcTn}
\end{center}
\end{figure}

In view of the following discussion, we concentrate on the anomalous features of $\kappa(T)$ in the vicinity of $T_N$. 
In Fig. \ref{KcTn}~a and b, we display the data for 
$\kappa(T)$ along the $c$-axis at temperatures 
between 10 and 13~K and for different values and
orientations of the magnetic field. In order to emphasize the change of the slope of $\kappa(T)$ at the transition,  we show the corresponding
temperature derivatives $\partial\kappa /\partial T$ in Fig. \ref{KcTn}~c and d.
The qualitative difference in the behavior for the two field orientations is obvious. 
While the transition, reflected in the sudden drop of $\partial\kappa /\partial 
T$ vs $T$ with decreasing $T$ occurs at 
the same temperature $T_{N}=11.4 {\rm ~K}$ for $H\parallel c \leq 60 {\rm kOe}$, 
the drop of  $\partial\kappa /\partial T$ vs. $T$ for $H \perp c \geq 20 {\rm kOe}$ is preceded by an initial increase, thus forming a narrow peak. With increasing $H$, the peak shifts to higher temperatures, obviously reflecting the anomalous $T_{N}(H)$-enhancement reported in the literature \cite{Lemmens01}. The absence of any variation of $T_N$ for  $H\parallel c$ has not been claimed before. 

Our data clearly demonstrate a feature in $\kappa(T)$ at $T_N$, at variance with the results  of Ref. \cite{Prester04}, where similar effects were observed  for the Cl-compound only. 
We suspect that the absence of an anomaly in $\kappa(T)$ at $T_N$ for the Br-compound in the data of Ref. \cite{Prester04} may be due to a dominating influence of defects in that Cu$_2$Te$_2$O$_5$Br$_2$ sample, masking the influence of the intrinsic scattering mechanisms that are related to the magnetic ordering.
Because of essentially the same features of $\kappa(T)$ at $T_N$ for the two compounds, we question the reasoning in Ref. \cite{Prester04} which suggests a 
drastic and intrinsic difference in the spin-phonon coupling between the two compounds.
The same crystal structure and the only slightly different size of the unit cell make the conjecture of Ref. \cite{Prester04} rather unlikely. 
Without presenting the taken data 
for the magnetic susceptibility $\chi(T)$ in the temperature  region between 2 and 300~K, we note that  
they exhibit all the characteristic features reported for $\chi(T)$ of the Cu$_2$Te$_2$O$_5$Br$_2$ single-crystal in Ref. \cite{Prester04}. These features include a  maximum of $\chi(T)$ at about 30~K, slightly higher $\chi$ values for $H\parallel c$ than $H \perp c$, the saturation to constant values  below about 5~K, and also the field-dependence of $\chi$ at low temperatures reported in Ref. \cite{Gros03}.

\section{Discussion}
In  magnetic insulators, energy may be transported in both the  crystal lattice and the spin system. 
In those cases where the approach of invoking excitations or quasiparticles is applicable, 
the total thermal conductivity can be represented as the sum of a phonon contribution $\kappa_{\rm ph}$ and a contribution of spin excitations (magnons, spinons etc.) $\kappa_s$. Each contribution $\kappa_i \propto C_i v_i \ell_i$ is given by the specific heat $C_i$ of the corresponding subsystem, the velocity of the related quasiparticles $v_i$ and their mean free path $\ell_i$. For each type of quasiparticle, the relaxation rate $\tau_i=\ell_i/v_i$ depends on the mutual interaction of the quasiparticles and the influence on their motion by various imperfections, such as point defects, dislocations, grain and domain boundaries etc. In many cases, $\tau_i^{-1}= \sum_j \tau_{i,j}^{-1}$, where $j$ corresponds to a particular type of scatterers. Among the various scattering processes, the spin-phonon interaction is of paramount importance in magnetic materials. Apart from influencing the magnitude and shape of $\kappa(T,H)$, the spin-phonon interaction provides, in a standard experimental arrangement of thermal-conductivity measurements, the necessary channel of heat transfer from the lattice to the spin system \cite{Sanders77}.

Since the magnon band in Cu$_2$Te$_2$O$_5$Br$_2$ is separated from the ground state by an energy gap $\Delta/k_B \approx 40 {\rm ~K}$, $\kappa_s(T)$ is expected to be negligibly small at $T \ll \Delta$ and to increase exponentially at $T \leq \Delta$. 
Any anisotropy of the spin interaction is expected to lead to an anisotropic increase in $\kappa_s(T)$. 
This expectation was confirmed in previous investigations of a number of quasi-1D and -2D magnetic systems, where
pronounced changes from a weakly temperature-dependent anisotropy (of phononic origin) at low temperatures to a strongly $T$-dependent anisotropy at higher temperatures were observed and interpreted as evidence for the onset of $\kappa_s$ (see discussion in Ref. \cite{Sologubenko01} and references therein). In Cu$_2$Te$_2$O$_5$Br$_2$, no change in the ratio $\kappa_a/\kappa_c$ is observed above the ordering transition (see Fig. \ref{KvsT_H0}), in spite of the anisotropy of the magnon spectrum along these directions \cite{Brenig03}.  This suggest that the spin contribution in Cu$_2$Te$_2$O$_5$Br$_2$ is negligibly small in comparison with the phonon contribution. 

The field-induced reduction of phonon transport is most pronounced and most sensitive to $H$ in the vicinity of the ordering transition. In magnetic materials, phonon-magnon scattering occurs via single-ion-lattice and magnetostrictive  interactions \cite{Kawasaki63,Stern65}. The intensity of the latter type of scattering is proportional to the magnetic specific heat $C_s(T)$ which exhibits an anomaly at  $T_N$. 
If  near the transition temperature, $C_s(T)$ is dominated by a 
discontinuity, usually leading to a peak-shaped anomaly, and if the phonons are predominantly scattered by the spin excitations, then a sharp dip in $\kappa(T)$ or, equivalently, a discontinuity in $\partial\kappa /\partial T$ vs $T$ is expected at $T_N$ \cite{Kawasaki63}. This is rarely the case for real materials where various $T$-dependent phonon-scattering mechanisms involving defects, boundaries, and the phonon-phonon interaction are stronger or at least of similar strength as the phonon-magnon scattering. If the anomaly in $C_s(T)$ is broadened for some reason, $\kappa(T)$ exhibits a broadly distributed reduction rather than the sharp dip mentioned above. 

\begin{figure}[t]
 \begin{center}
  \epsfxsize=0.9\columnwidth \epsfbox {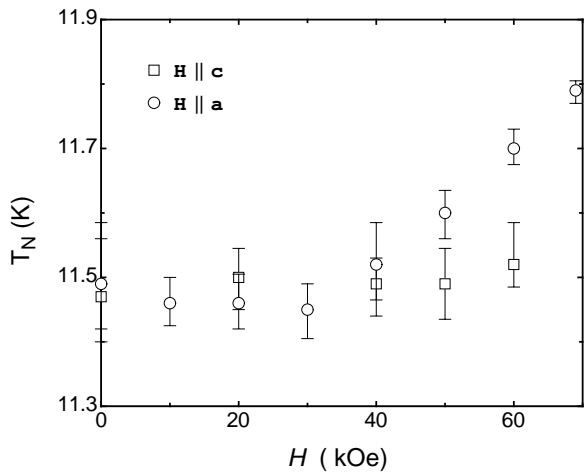}
   \caption{
  The temperature of the ordering transition as a function of magnetic field. 
  }
\label{Tn}
\end{center}
\end{figure}

Specific heat data by Lemmens {\em et al.} \cite{Lemmens01}, taken on a powder sample of  Cu$_2$Te$_2$O$_5$Br$_2$, reveal, upon the application of an external magnetic field, the growth and shift of the broad peak of $C(T)$ at $T_N$ to higher temperatures. This correlates with the enhanced reduction of $\kappa(T)$ and the shift of the related anomaly to higher temperatures, which we observe in our experiments for $H \parallel a$, see Fig. \ref{KcTn}~(a,c). 
It is to be noted, however, that no such effect is observed for $H \parallel c$. The transition temperature $T_N(H)$, calculated from the $\partial\kappa /\partial T$ vs $T$ data, is shown in Fig. \ref{Tn}.
As mentioned,  $T_N$ increases with $H\parallel a$, in agreement with the quoted earlier observations \cite{Lemmens01}; the critical temperature does not change in the investigated field region for $H \parallel c$, however.

An increasing $T_N$ with increasing field cannot simply be explained by the classical theory of antiferromagnetism.  Nonetheless, such behavior has been observed for several AFM compounds \cite{Oliveira79,Clark75,Hijmans78,Honda98,Honda01}. 
It was argued that this behavior is related to the fact that, generally, a lower spin dimensionality leads to higher values of the critical field.
A magnetic field oriented along particular crystallographic directions may lead to a  reduction of the effective spin dimensionality \cite{Fisher74,Fisher75}.
However, if the magnetic field is oriented in such a way that it does not change the spin dimensionality, such as along the easy axis of an Ising model system or perpendicular to the easy plane of a planar model, an MF theory-consistent reduction of $T_N$ with increasing field is expected. 
For 3D AFM spin systems, the dimensionality-driven enhancement of $T_N$ is, at most, of the order of 0.1\% \cite{Oliveira79}, but quantum effects in spin compounds containing structural elements with lower dimensionality, particularly in spin-chain compounds \cite{imry75}, lead to much stronger field-induced changes of $T_N$ \cite{Clark75,Hijmans78,Honda98,Honda01}.    
If the same type of arguments is valid for the anisotropic shifts of $T_N(H)$ observed in this work for tetragonal Cu$_2$Te$_2$O$_5$Br$_2$, it may be concluded that the magnetic order is characterized by an easy-axis moment orientation along the $c$-axis. 

This conclusion is in agreement with the results of  calculations of Jensen {\em et al.} \cite{Jensen03}, which include a Dzyaloshinskii-Moriya type anisotropy term in the Hamiltonian describing a  model of dimerized interacting tetrahedra \cite{Gros03}. 
In relation with the Br-compound, the model calculations predict 
an AFM ordered state below $T_N$ with staggered moments aligned along the $c$-axis. The calculations also predict $T_N$ to increase with $H \perp c$ for all values of $H$, consistent with our result, but also a weak initial decrease of $T_N(H)$ for $H \parallel c$, intercepted by a spin-flop transition at about 37~kOe. The latter transition is not reflected in our data, but the calculated value of the spin-flop field is parameter-dependent. It is certainly not inconceivable that in reality, this field is higher than the calculated value of 37~kOe. An increase of $T_N$ with the mentioned field configuration is also consistent with another analysis of a model of coupled spin tetrahedra taking into account a Dzyaloshinskii-Moriya-type interaction \cite{Kotov04cm}. These calculations also predict a decrease of $T_N(H)$ at almost the same rate, if a field is applied in the $c$-direction. This expectation is not supported by our experiments, however.

\section{Summary}
In this work the thermal conductivity of the spin-tetrahedral compound Cu$_2$Te$_2$O$_5$Br$_2$ has been studied. The results clearly indicate that phonons dominate the heat transport in this compound. A feature in $\kappa(T)$ at $T_N$, close to 11.4~K, is associated with a magnetic ordering transition. The transition temperature and the amplitude of the associated $\kappa(T)$ anomaly are affected by external magnetic fields only if they are oriented along the $a$-axis. This $T_N(H)$ anisotropy is qualitatively consistent with recent theoretical predictions.  

\acknowledgments
This work was financially supported in part by
the Schweizerische Nationalfonds zur F\"{o}rderung der Wissenschaftlichen
Forschung.


\end{document}